\documentclass[aps,prd,twocolumn,showpacs,superscriptaddress,groupedaddress,eqnleft]{revtex4}  
\usepackage[utf8]{inputenc}
\usepackage[T1]{fontenc}
\usepackage[english]{babel}
\usepackage{amsfonts,amsmath,amssymb,bbm,slashed}
\usepackage[amssymb,Gray]{SIunits}
\usepackage{url}
\usepackage{xspace}
\newcommand{\rmi}{\ensuremath{\mathrm{i}}}
\newcommand{\rme}{\ensuremath{\mathrm{e}}}
\newcommand{\real}{\ensuremath{\mathrm{Re}}}
\newcommand{\imag}{\ensuremath{\mathrm{Im}}}
\newcommand{\erf}{\ensuremath{\mathrm{erf}}}
\newcommand{\abs}[1]{\ensuremath{\left\vert #1 \right\vert}}

\newcommand{\order}[1]{\mathcal{O}(#1)}
\newcommand{\bleq}{\ensuremath{\mathrel{\phantom{=}}}}
\newcommand{\nnl}{\nonumber\\}

\newcommand{\tmatrix}[4]{\left(\begin{array}{cc}#1&#2\\#3&#4\end{array}\right)}
\newcommand{\bra}[1]{\langle #1 \hspace{-2pt} \mid}
\newcommand{\ket}[1]{\mid \hspace{-1pt} #1 \rangle}

\newcommand{\D}{\mathrm{d}}
\newcommand{\laplace}{\nabla^2}

\newcommand{\sne}{Schr{\"o}\-din\-ger--New\-ton equation\xspace}
\newcommand{\schr}{Schr{\"o}\-din\-ger\xspace}

\renewcommand{\vec}[1]{\mathrm{\mathbf{#1}}}
\newcommand{\tvector}[2]{\left(\begin{array}{c}#1\\#2\end{array}\right)}

\begin{document}

\title{Newtonian Self-Gravitation in the Neutral Meson System}
\author{Andr\'e Gro{\ss}ardt}
\affiliation{Department of Physics, University of Trieste, 34151 Miramare-Trieste, Italy}
\affiliation{Istituto Nazionale di Fisica Nucleare, Sezione di Trieste, Via Valerio 2,
34127 Trieste, Italy}
\author{Beatrix C. Hiesmayr}
\affiliation{Faculty of Physics, University of Vienna, Boltzmanngasse 5,
1090 Vienna, Austria}
\date{\today}

\begin{abstract}
We derive the effect of the Schrödinger--Newton equation, which can be considered as a non-relativistic limit of
classical gravity, for a composite quantum system in the regime of high energies. Such meson-antimeson systems
exhibit very unique properties, e.\,g. distinct masses due to strong and electroweak interactions. We find
conceptually different physical scenarios due to lacking of a clear physical guiding principle which mass is the
relevant one and due to the fact that it is not clear how the flavor wave-function relates to the spatial wave-function.
There seems to be no principal contradiction. However, a nonlinear extension of the Schr\"odinger equation in
this manner
strongly depends on the relation between the flavor wave-function and spatial wave-function and its particular shape.
In opposition to the Continuous Spontaneous Localization collapse models we find a change in the oscillating behavior
and not in the damping of the flavor oscillation.
\end{abstract}

\maketitle

\section{Introduction}

The search for a theory that consistently combines quantum theory and gravitation is certainly
one of the bigger challenges of contemporary theoretical physics.
While most physicists believe that---whatever the correct quantum theory of gravity is---in the low-energy
limit gravity can be described by a perturbative quantum field theory, in full analogy to the low-energy limit of
Quantum Electrodynamics, there is no experimental evidence, to date, that rules out a theory in which
gravity remains unquantized, even at the fundamental level. This idea has been raised by many
before~\cite{Rosenfeld:1963,Kibble:1981,Mattingly:2005,Carlip:2008}, and a behavior of gravity
at the quantum level that is different from what one would expect by a naive perturbative quantization of the
gravitational field is also discussed as a possible solution of the quantum measurement
problem~\cite{Diosi:1984,Diosi:1987,Diosi:1989,Diosi:2007,Penrose:1996,Penrose:1998,Penrose:2014,Adler:2014}.
In the context of non-relativistic quantum mechanics, the problem is basically condensed to the question how
quantum matter sources the gravitational field.

One hypothesis that has been brought into the debate~\cite{Carlip:2008,Giulini:2012,Giulini:2014,Bahrami:2014}
is that the gravitational interaction for non-relativistic quantum matter is described by a
nonlinear extension of the \schr equation, the \sne
\begin{eqnarray}
\label{eqn:sn}
\lefteqn{\rmi \hbar \partial_t\, \psi(t,\vec r)\;=}\nonumber\\ 
&& \Bigg( -\frac{\hbar^2}{2 m} \laplace
- G m^2 \int \D^3 \vec r' \, \frac{\abs{\psi(t,\vec r')}^2}{\abs{\vec r - \vec r'}}
\Bigg) \psi(t,\vec r) \,,
\end{eqnarray}
originally proposed as a model for the localization of macroscopic quantum objects~\cite{Diosi:1984,Penrose:1998}.
The intuition behind such an equation is that the absolute-value squared of the wave-function corresponds
to a mass density sourcing a Newtonian gravitational potential~\footnote{While such a mass-density interpretation
would be incompatible with the instantaneous collapse and the perception of the wave-function as a pure probability
density in the Copenhagen interpretation, a dynamical
state-reduction as in collapse models~\cite{Bassi:2013} allows for such an interpretation.}.
The equation can also be shown to follow naturally as the non-relativistic limit of a semi-classical theory
of gravity, i.\,e. a theory in which the gravitational field stays classical even at the fundamental level and quantum
matter is coupled by the semi-classical Einstein equations
\begin{equation}
\label{eqn:sce}
R_{\mu \nu} + \frac{1}{2} g_{\mu \nu} R = \frac{8 \pi G}{c^4} \,
\bra{\Psi} \hat{T}_{\mu \nu} \ket{\Psi} \,,
\end{equation}
where $\hat{T}_{\mu \nu}$ is the energy-momentum operator and the expectation value is taken in some
quantum state~\cite{Bahrami:2014}.
One particularly charming aspect of the \sne is that it most likely can be experimentally tested in the foreseeable
future, e.\,g., with large molecules~\cite{Giulini:2011,Giulini:2013,Yang:2013} or with crystalline
nanospheres~\cite{Colin:2014}.

In this letter we want to consider neutral meson-antimeson systems that are typically produced at accelerator facilities.
In particular we focus for the sake of simplicity on the neutral K-meson system, also dubbed kaons, however, all
considerations hold for all non-relativistic meson systems. These massive systems, that can also be produced even in entangled pairs, have
shown to be a unique laboratory for precisions measurements of particle properties and fundamental principles in particle
physics (e.\,g. discrete symmetries) as well as for testing fundamental principles in quantum physics such as
superposition and entanglement (for an overview see e.\,g. ref.~\cite{Amelino-Camelia:2010}). For example, a violation of
Bell's inequality that is only due to the breaking of a discrete symmetry resulting in a tiny difference between matter
and antimatter properties has been discovered~\cite{Hiesmayr:2012}. Or due to the existence of two distinct measurement
procedures, a special feature of kaons, the very working of a quantum
eraser~\cite{Bramon:2004,Bramon:2004a} or Heisenbergs principle~\cite{DiDomenico:2012} can in a novel way be
demonstrated. Proposals how to test decoherence effects have been developed~\cite{Bertlmann:1999,Hiesmayr:2001} and put
to experimental tests~\cite{Ambrosino:2006,DiDomenico:2009,Go:2007}. Models testing for Lorentz-symmetry violations or
assuming intrinsic violations of the CPT symmetry induced by quantum gravity~\cite{Bernabeu:2006} have been put to test
for K-mesons~\cite{Babusci:2014,Ambrosino:2006}. Recently, also the prediction of collapse models were
computed~\cite{Bahrami:2013,Donadi:2013}.

Despite the fact that these meson systems are elementary particle systems, for which one would expect that they must
be theoretically treated with the tools of relativistic quantum field theories, the formalism of non-relativistic
quantum mechanics turns out to provide good predictions for almost all interesting effects in these systems.

At first sight neutral kaons, composite systems of a quark and an anti-quark, seem not to be good candidates to test for
gravitational effects since the mass is very low, approximately half of a proton mass, however, the unique properties of
these meson-antimeson systems---as witnessed by the above literature---makes it an interesting case to see whether
conceptual contradictions can be derived. Let us here quote the famous Feynman lectures~\cite{Feynman:1965}, where
Feynman writes after introducing K-mesons:

\begin{center}\textit{``If there is any place where we have a chance to test the main
principles of quantum mechanics in the purest way —does the
superposition of amplitudes work or doesn’t it?— this is it.''}
\end{center}

The superposition of these two different mass eigenstates due to weak interaction exhibiting oscillations of the
eigenstates of the strong interaction have been proven by now at many accelerator facilities. Moreover, since 1964 the
unexpected breaking of the CP symmetry (C\dots charge conjugation, P\dots parity), a tiny difference between matter and
antimatter properties, was discovered.

Taking the point of view that the \sne\ correctly describes the coupling of quantum matter to
gravity---keeping in mind that as a mere hypothesis it could be experimentally falsified at any time---one
may immediately ask:\\
\begin{center}
\textit{
Do both eigenstates of the mass-Hamiltonian couple independently to the gravitational field?\\ Or is only the rest mass of the neutral K-meson the one relevant for any gravitational effect?}
\end{center}

In this contribution we analyze in detail which options to include a \schr--Newton interaction in the
neutral K-meson system are conceptually possible---if any---and which effects they may have on
the flavor oscillations.
We will briefly review the properties of the neutral kaon system in the second section.
There we raise the important question of how the spatial wave-function should be treated in the case of neutral mesons.
We then discuss different ways to implement the features of the neutral kaon into the \sne in the third
section, accounting for the dependence on the right description of the spatial wave-function.
In the fourth section we compare these results to the previously obtained results for the
Continuous Spontaneous Localization (CSL) collapse model.
Finally, we discuss the results and draw our conclusions.

\section{The neutral kaon system}

Via strong interactions one has to distinguish between two different eigenstates labeled by the strangeness number $S$,
the kaon state $\ket{K^0}$ ($S=1$) and the
anti-kaon $\ket{\bar{K}^0}$ ($S=-1$). Neutral kaons decay via the weak interaction leading to the following
non-Hermitian Hamiltonian
\begin{align}
H &= \left(\begin{array}{cc}\langle K^0|H^{(|\Delta S|=0)}|K^0\rangle&\langle \bar K^0|H^{(|\Delta S|=2)}|K^0\rangle\\
\langle K^0|H^{(|\Delta S|=2)}|\bar K^0\rangle & \langle \bar K^0|H^{(|\Delta S|=0)}|\bar
K^0\rangle\end{array}\right) \nnl
&=M-\frac{\rmi}{2} {\bf \Gamma}
\end{align}
where both the mass matrix $M$ and decay-matrix ${\bf \Gamma}$ are chosen to be Hermitian. $H^{(|\Delta S|=0)}$ describes
the processes which conserve the strangeness number $S$ and $H^{(|\Delta S|=2)}$ those which differ by two. The states
diagonalising this Hamiltonian are denoted as mass eigenstates, namely the short- and long-lived states $\ket{K_S}$ and
$\ket{K_L}$. If we assume $CPT$ conservation ($T$\dots time reversal), the two diagonal elements of $M$ have to be equal
and have to correspond to the rest mass $m_K$. Analogously, the two diagonal elements of ${\bf \Gamma}$ have to be equal
and to correspond to the total decay width $\Gamma$ of $K^0, \bar K^0$.

The complex eigenvalues of the Hamiltonian $H$ are derived to
\begin{align} \label{eqn:complex_eigenval}
\lambda_{S/L} &= m_{S/L}-\frac{\rmi}{2} \Gamma_{S/L} \nnl
&= m_K-\frac{\rmi}{2} \Gamma\mp\sqrt{(M_{12}-\frac{\rmi}{2}
{\bf \Gamma}_{12})(M_{12}^*-\frac{\rmi}{2} {\bf \Gamma}_{12}^*)}
\end{align}
and, consequently the mass difference $\Delta m:=m_L-m_S$ and the decay width difference
$\Delta \Gamma:= \Gamma_L-\Gamma_S$ are given by
\begin{eqnarray}
\Delta m &=& 2\real\{\sqrt{(M_{12}-\frac{\rmi}{2} {\bf \Gamma}_{12})(M_{12}^*-\frac{\rmi}{2}
{\bf \Gamma}_{12}^*)}\}\nonumber\\
\Delta \Gamma &=& -4 \imag\{\sqrt{(M_{12}-\frac{\rmi}{2} {\bf \Gamma}_{12})(M_{12}^*-\frac{\rmi}{2}
{\bf \Gamma}_{12}^*)}\}\;.
\end{eqnarray}
The mass difference $\Delta m$ and the two decay widths $\Gamma_S,\Gamma_L$ have been measured for all neutral meson
systems~\cite{Nakamura:2010}, however, only for neutral K-mesons the two decay widths differ greatly.

Thus the time evolution of the mass-Hamilton eigenstates is given by\footnote{In this section we use natural units,
$\hbar = c = 1$.}
\begin{align}
 \ket{K_S(t)} &= \mathrm{e}^{-\rmi m_S t} \mathrm{e}^{- \frac{\Gamma_S}{2} t} \ket{K_S(t=0)} \nnl
 \ket{K_L(t)} &= \mathrm{e}^{-\rmi m_L t} \mathrm{e}^{- \frac{\Gamma_L}{2} t} \ket{K_L(t=0)} \,,
\end{align}
preserving their identity in time. The mass eigenstates states are connected via the following basis transformation
\begin{align}
 \ket{K_S} &= \frac{(1+\varepsilon) \ket{K^0} - (1-\varepsilon)\ket{{\bar K}^0}}{\sqrt{2(1+|\varepsilon|^2)}} \nnl
 \ket{K_L} &= \frac{(1+\varepsilon)\ket{K^0} + (1-\varepsilon)\ket{{\bar K}^0}}{\sqrt{2(1+|\varepsilon|^2)}} \,.
\end{align}
where $\varepsilon$ is the CP violating parameter that equals in a conventional phase choice to
\begin{eqnarray}
\varepsilon &=& \frac{(M_{12}-\frac{\rmi}{2} {\bf \Gamma}_{12})-(M_{12}^*-\frac{\rmi}{2}
{\bf \Gamma}_{12}^*)}{(M_{12}-\frac{\rmi}{2} {\bf \Gamma}_{12})+(M_{12}^*-\frac{\rmi}{2} {\bf \Gamma}_{12}^*)}\;.
\end{eqnarray}
To understand the difference between the dynamical parameters $\Delta m,\Delta\Gamma$ and the CP violating parameter
$\varepsilon$ let us introduce two complex numbers $X,Y$ for which it is straightforwardly to show that if we define
\begin{eqnarray}
\frac{Y}{X}=M_{12}-\frac{\rmi}{2}{\bf \Gamma}_{12}
\end{eqnarray}
the equation $X\cdot Y =M_{12}^*-\frac{\rmi}{2}{\bf \Gamma}_{12}^*$ holds. With that we find (up to non-physical sign
changes)
\begin{eqnarray}
\Delta m &=&2 \real\{Y\}\nonumber\\
\Delta \Gamma &=&-4 \imag\{Y\}\nonumber\\
\varepsilon&=&\frac{1-X^2}{1+X^2}\;.
\end{eqnarray}
Obviously the values $\Delta m,\Delta\Gamma$ are independent of $\varepsilon$ in the sense that the value of $X$ does not
influence the value of these dynamical parameters, however, the time evolution does depend on all three parameters as we
show explicitly in the following.

The probabilities of finding a $K^0$ or a $\bar K^0$ after a certain time $t$ if a state $|K^0\rangle$ was produced at
time $t=0$ is consequently given by
\begin{align}\label{eqn:flavor-oscillations}
P(K^0 t;|K^0|) &= |\langle K^0|K^0(t)\rangle|^2 \nnl
&= \frac{1}{4}\left(\rme^{-\Gamma_S t}+ \rme^{-\Gamma_L t}+2
\cos(\Delta m\, t)\cdot \rme^{-\Gamma t}\right)\\
P(\bar K^0 t;|K^0|) &= |\langle \bar K^0|K^0(t)\rangle|^2 \nnl
&= \frac{|1-\varepsilon|^2}{4 |1+\varepsilon|^2}\Big(\rme^{-\Gamma_S t}+ \rme^{-\Gamma_L t} \nnl
&\bleq -2 \cos(\Delta m\, t)\cdot \rme^{-\Gamma t}\Big)\;.
\end{align}
Taking the difference we derive the time dependent asymmetry
\begin{eqnarray}
\frac{P(K^0 t;|K^0|)-P(\bar K^0 t;|K^0|)}{P(K^0 t;|K^0|)+P(\bar K^0 t;|K^0|)}&=&\frac{\frac{2 \real\{\varepsilon\}}{1+|\varepsilon|^2}+\frac{\cos(\Delta m \,t)}{\cosh(\frac{\Delta\Gamma}{2} t)}}{1+\frac{2 \real\{\varepsilon\}}{1+|\varepsilon|^2}\frac{\cos(\Delta m \,t)}{\cosh(\frac{\Delta\Gamma}{2} t)}}\;,\nonumber\\
\end{eqnarray}
where for short times the oscillation is visible whereas in the long time limit the $CP$ violation can be measured. In
summary one observes a damped oscillation due to $\Delta m$ and the decay constants where the mass $m_K$ does not enter
and for the huge time regime the difference of both  probabilities reveals the tiny CP violation, namely
$\frac{2 \real\{\varepsilon\}}{1+|\varepsilon|^2}\approx 10^{-3}$.

Let us remark on the non-relativistic treatment of K-mesons as composite systems. Ordinarily, nonrelativistic
systems have excitation energies that
are small compared to the component masses. For mesons systems, however, these energies are comparable to the quark
masses of these models. The complexity of QCD forces one to resort to approximate models, so called bag models (see,
e.\,g., the review article~\cite{DeTar:1983}). These bag models divide space into two regions, the interior of the bag in
which the quarks have very small (current) masses and feel only weak forces and the exterior in which the quarks are not
allowed to propagate having a different (lower) vacuum energy. In the above presented phenomenology, mesons are treated as
a single entity, in very good agreement to all current experiments. Relativistic effects---such as the speed of the
mesons---do not alter the physics in the flavor space, thus we do not consider any particular relativistic effects in the
following treatment.

Certainly, we can be interested in the kinematics of the neutral meson system, then the relevant Hamiltonian would be
\begin{equation}
H_\text{kin} = m_{\textrm{inert}} c^2 -\frac{\hbar^2}{2 m_{\textrm{inert}}} \laplace \,.
\end{equation}
The inertial mass $m_{\textrm{inert}}$ could be considered as that of the composite K-meson, i.\,e.  $m_K$, or one may
assume that each mass-energy eigenstate $K_{S/L}$ exhibits a different kinetic/spatial wave-function due to its
different decaying property, namely $m_{S/L}$. Note that $m_{S/L}$ alone has also contributions of $m_K$. Differently
stated, considering the spatial wave-function of mesons:

\begin{center}\textit{Do we have to consider only one unique wave-function or do we have to handle it as a two--state
system?}\end{center}

We will differentiate between those two scenarios in the following.

Before we proceed let us comment on the validity of a non-relativistic treatment of the K-meson via the
Schr\"odinger equation. The first point to mention is that we are interested in space-dependent probabilities
measuring the strangeness content, i.\,e. probabilities that a neutral kaon, having propagated a certain macroscopic
distance, decays semileptonic or is forced by a matter block to reveal over the subsequent reaction the strangeness
property. Flavor oscillation, i.\,e. the probabilities of observing a particle or anti-particle state at a certain
position/time, is the phenomenon one is interested in. This distance is usually converted into a proper time
(via $\tau \approx\frac{x}{v}$). Here the relativistic effects matter but of course in a trivial way.  Given the
Hamiltonian above, the Schr\"odinger equation is the appropriate non-relativistic limit of either the
Klein--Gordon equation or Dirac equation keeping the rest mass energy. There have been numerous approaches, how
the spatial wave-function of a
strangeness state---that has to be a coherent state composed of a short-lived and long-lived state---has to be treated
in an unambiguous way~\cite{Ancochea:1996,Lipkin:1995,Burkhardt:2003,Okun:2003} (and for the neutrino oscillations, see
e.\,g.~\cite{Kim:1993}). The main disagreement is whether the energies and/or momenta of the two different
mass eigenstates remain unchanged. However, the final probabilities, i.\,e. equation~\eqref{eqn:flavor-oscillations},
are---within good approximations---always the same, that are also those well tested in various experiments. The authors
of~\cite{Ancochea:1996} apply a different view by defining the probabilities as position measurements with averages over
times. In summary, the key observation is that experiments are only sensitive to the mass difference and not sum of mass
nor energy, thus experiments are not restricted to favor one approach over the other one and only conceptual arguments
apply.  Modification of the standard quantum approach such as the here discussed
Schr\"odinger-Newton equation or collapse models~\cite{Bahrami:2013,Donadi:2013}, however, require an explicit modelling
of the relation between the spatial wave function and the flavor space and, therefore, offer a unique laboratory to
enlighten the subtle interplay given rise to particle-antiparticle mixing.

\section{\sne for the neutral kaon}

The \sne is based on the assumption that the wave-function sources a gravitational field as if the total mass of
the particle would be smeared with the spatial probability density $\abs{\psi}^2$. If the \sne is applied to the
kaon system this raises two questions:
\begin{itemize}
 \item Which is the right mass that acts as the source of the gravitational field?
 \item How to describe the spatial wave-function of the neutral kaon?
\end{itemize}
And particularly, does this lead to any inconsistencies or unexpected effects that could render the
\sne\ an ill-defined model for describing these systems, or on the contrary lead to observable effects?

Due to Newton the mass entering into the kinetic term of the Hamiltonian is the inertial mass. In the famous experiment
with neutrons the authors of ref.~\cite{Colella:1975} demonstrated that non-relativistic quantum matter can couple to the
gravitational field in the following way:
\begin{equation}
\label{eqn:se-grav}
\rmi \hbar\, \partial_t \psi(t,\vec r) = \left( -\frac{\hbar^2}{2 m_\text{inert}} \laplace
+ m_\text{grav}^\text{passive}\, \Phi_\text{grav} \right) \psi(t,\vec r) \,.
\end{equation}
Here $\Phi_\text{grav}$ is the gravitational potential, and
$m_\text{grav}^\text{passive}$ is the passive gravitational mass, i.\,e. the coupling of matter to the
gravitational field. If the weak equivalence principle holds, these two masses must be equal:
$m_\text{grav}^\text{passive} = m_\text{inert}$. Let us remark here that so far in all situations that have been put to
test by experiments,
$\Phi_\text{grav}$ belongs to an \emph{external} gravitational field.

In the framework of the \sne, however, $\Phi_\text{grav}$ yields the gravitational self-interaction.
With the assumptions underlying the \sne the gravitational potential satisfies the Poisson equation
\begin{equation} \label{eqn:poisson}
 \laplace \Phi_\text{grav} = 4 \pi G \, m_\text{grav}^\text{active} \, \abs{\psi(t,\vec r)}^2 \,.
\end{equation}
The mass entering here as the source of the gravitational field is referred to as the active gravitational mass.
Note that neither the equivalence principle nor any other fundamental principle of physics demands that this mass
is equal to the inertial mass or passive gravitational mass.\footnote{In the classical limit, of course,
the active and passive gravitational mass are equal due to Newton's third law.}

Interestingly, equation~\eqref{eqn:poisson} can be shown to follow from the semi-classical Einstein
equations~\eqref{eqn:sce} as derived in refs.~\cite{Giulini:2013,Bahrami:2014}. The mass density on the right-hand side
is then
given by the expectation value of the non-relativistic limit of the energy-momentum operator,
\begin{equation}
 m_\text{grav}^\text{active} \, \abs{\psi}^2  = \frac{1}{c^2} \,\bra{\psi} \hat{T}_{00} \ket{\psi} \,.
\end{equation}
Following this logic, from a quantum field theoretical point of view the mass $m_\text{grav}^\text{active}$
would be the one appearing in the mass term of the field Lagrangian (after renormalization).
But the kaon is a composite system and its mass is mainly binding energy of quarks.

Usually, one assumes that all three masses, inertial as well as active and passive gravitational mass,
correspond to $m_K = \unit{(497.614\pm0.024)}{\mega\electronvolt}\,c^{-2}$,  the measurable invariant mass of the neutral
kaon.
Whereas the flavor eigenstates $|K^0\rangle,|\bar K^0\rangle$ have equal mass (assuming $CPT$ symmetry) with the value
$m_K$, the long-lived and short-lived eigenstates $|K_S\rangle,|K_L\rangle$ of the strong and weak interaction
Hamiltonian manifest a small mass difference $\Delta m=(3.483\pm0.006){\mega\electronvolt}\,c^{-2}$ that is at another
energy scale, as discussed in the previous section.

\begin{center}\textit{So, does this mass difference
$\Delta m$ show up in the \sne, if so, for which of the three masses does it show up, and what would be the
consequences?}\end{center}

As indicated in the previous section, similar conceptual
questions arise concerning the
spatial wave-function $\psi(\vec r)$ of the non-relativistic neutral kaon.
Again, one could simply consider a Schr\"odinger equation with one unique spatial wave-function that evolves with
the invariant mass, $m_K$, in the kinetic term. But since the states $\ket{K_S}$ and $\ket{K_L}$ diagonalize the
Hamiltonian, there should in principle be different wave-functions $\psi_S(\vec r)$ and $\psi_L(\vec r)$
evolving with the masses $m_S$ and $m_L$, respectively.
Because of the difference in the free spreading of the wave-function due to the mass difference one would then
expect additional flavor oscillations in space.

Therefore, we will in the following distinguish two scenarios and discuss their implications:\\
\\
\textbf{Scenario 1: Unique spatial wave-function}\\
\\
Let us first assume that we can treat $\ket{K_S}$ and $\ket{K_L}$ as if they would mix only in flavor space,
while they are described by one unique spatial wave-function. The total wave-function is then
\begin{eqnarray}
 \lefteqn{\psi_\text{flavor} \otimes \psi_\text{space} \;=}\nonumber\\
 && (\alpha(t)\; \psi_\text{S,flavor} + \beta(t)\; \psi_\text{L,flavor})
\otimes \psi_\text{space} \,,
\end{eqnarray}
where $\psi_\text{space}$ denotes the spatial wave-function and $\psi_\text{flavor}$ the flavor part. $\psi_\text{S,flavor}$ and $\psi_\text{L,flavor}$ denote a short-lived and long-lived kaon, respectively.
$\psi_\text{space}$ satisfies the \sne for the mass $m^\text{inert} = m_\text{grav}^\text{passive} = m_K$.
The time-dependent coefficients $\alpha$ and $\beta$ represent the decomposition of the current kaon state in the
basis $\{\ket{K_S},\ket{K_L}\}$. They are time dependent due to the flavor oscillations, i.\,e.
\begin{equation}
\alpha(t) = \alpha_0 \, \rme^{-\rmi \, \Delta m \, t} \,, \quad
\beta(t) = \beta_0 \, \rme^{-\rmi \, \Delta m \, t} \,.
\end{equation}

If the active gravitational mass does not depend on the flavor part, the equation can be separated and
the spatial wave-function will simply satisfy the \sne for the kaon mass $m_K$, independent of its composition
of $\ket{K_S}$ and $\ket{K_L}$.

If, on the other hand, $m_\text{grav}^\text{active}$ does depend on the flavor part, the most naive ansatz would
be $m_\text{grav}^\text{active} = \abs{\alpha(t)}^2 m_S + \abs{\beta(t)}^2 m_L $.
One then obtains the \sne
\begin{align}
\label{eqn:sn_one_wf_two_masses}
\rmi \hbar \partial_t \psi_\text{space}(t,\vec r) &= \Bigg( H_\text{kin}(m_K) \nnl
&\bleq\hspace{-70pt} - G m_K m_S\,  \abs{\alpha(t)}^2 \, \int \D^3 \vec r' \,
\frac{\abs{\psi_\text{space}(t,\vec r')}^2}{\abs{\vec r - \vec r'}} \nnl
&\bleq\hspace{-70pt} - G m_K m_L \, \abs{\beta(t)}^2 \, \int \D^3 \vec r' \,
\frac{\abs{\psi_\text{space}(t,\vec r')}^2}{\abs{\vec r - \vec r'}}
\Bigg) \psi_\text{space}(t,\vec r) \,.
\end{align}
Since the mass difference $\Delta m = m_L - m_S$ is small, the effect is only a tiny modification of the
already unmeasurable gravitational self-interaction.\\
\\
\textbf{Scenario 2: Different spatial wave-functions}\\
\\
Now let us consider the case assuming that $\ket{K_S}$ and $\ket{K_L}$ have different wave-functions
also in space. Therefore, the total wave-function is given by
\begin{equation}
\begin{split}
 \psi_\text{flavor} \otimes \psi_\text{space} = \alpha(t)\; \psi_\text{S,flavor} \otimes \psi_\text{S,space}
\\ + \beta(t)\; \psi_\text{L,flavor} \otimes \psi_\text{L,space} \,.
\end{split}
\end{equation}
Each of the spatial wave-functions will contribute to the total gravitational potential, and both wave-functions
will see this same gravitational potential.
Therefore, we get the following two \schr--Newton equations:
\begin{subequations}\label{eqn:sn-different-wavefunctions}\begin{align}
 \rmi \hbar \partial_t \psi_S(t, \vec r) &= \Bigg( H_\text{kin}(m_S) \nnl
&\bleq\hspace{-50pt} - G m_S^2\,  \abs{\alpha(t)}^2 \, \int \D^3 \vec r' \,
\frac{\abs{\psi_S(t,\vec r')}^2}{\abs{\vec r - \vec r'}} \nnl
&\bleq\hspace{-50pt} - G m_S m_L\,  \abs{\beta(t)}^2 \, \int \D^3 \vec r' \,
\frac{\abs{\psi_L(t,\vec r')}^2}{\abs{\vec r - \vec r'}}
\Bigg) \psi_S(t, \vec r) \\
\rmi \hbar \partial_t \psi_L(t, \vec r) &= \Bigg( H_\text{kin}(m_L) \nnl
&\bleq\hspace{-50pt} - G m_S m_L\,  \abs{\alpha(t)}^2 \, \int \D^3 \vec r' \,
\frac{\abs{\psi_S(t,\vec r')}^2}{\abs{\vec r - \vec r'}} \nnl
&\bleq\hspace{-50pt} - G m_L^2\,  \abs{\beta(t)}^2 \, \int \D^3 \vec r' \,
\frac{\abs{\psi_L(t,\vec r')}^2}{\abs{\vec r - \vec r'}}
\Bigg) \psi_L(t, \vec r) \,,
\end{align}\end{subequations}
where we write $\psi_{S,L}$ for the spatial wave-functions of the short and long lived contribution, respectively.

As discussed in the previous section, both scenarios are compatible with what has been experimentally tested so far,
although there seems to be a slight preference for this latter scenario in the literature.
Note that this second scenario, where $K_S$ and $K_L$ evolve with different spatial wave-functions, is also the only one
that is compatible with a derivation of the \sne from a doublet-state formalism, cf. appendix~\ref{app:doublet}.

\subsection{Resulting wave-function dynamics}

In experimental situations, the kaon is usually not well-localized. Therefore, the wave-function is usually
assumed to be a plane wave, in very good agreement with the experiment. However, to determine the effect of
the \sne, the localization of the wave-function must be taken into account. We will therefore model it by a
spherically symmetric Gaussian:
\begin{equation}\begin{split} \label{eqn:free-gaussian}
 \psi^f(t,r;m,a) = \left(\pi a^2\right)^{-3/4} \left(1+\frac{\rmi \hbar t}{m a^2}\right)^{-3/2}\\
\times \exp \left(-\frac{r^2}{2 a^2 \left(1+\frac{\rmi \hbar t}{m a^2}\right)} \right) \,.\end{split}
\end{equation}
This is the solution of the free \schr equation, where the width, $a$, will be assumed to be large. In general, the
\schr--Newton dynamics disturb the Gaussian shape of the wave-function. Since the gravitational
interaction is very weak due to the large value of $a$ and the small mass $m$, we will approximate the
wave-function appearing as the mass density in the gravitational potential by the free
solution~\eqref{eqn:free-gaussian}. The approximated gravitational potential is then
$\Phi_\text{grav} = -G\,m_\text{grav}^\text{active} f(t,r;m,a)$ with
\begin{align}
f(t,r;m,a) &= \int \D^3 \vec r' \, \frac{\abs{\psi^f(t,\vec r';m,a)}^2}{\abs{\vec r - \vec r'}} \nnl
&= \frac{1}{r} \erf\left[\frac{r}{a} \left(1+\frac{\hbar^2 t^2}{m^2 a^4}\right)^{-1/2} \right]
\end{align}

The function $f$ could now be expanded in terms of the mass, yielding
\begin{equation} \label{eqn:mass-expansion}
 f(t,r;m,a) = \frac{2 a m}{\sqrt{\pi} \hbar t} + \order{m^3} \,,
\end{equation}
or in terms of time, yielding
\begin{eqnarray} \label{eqn:time-expansion}
\lefteqn{f(t,r;m,a)\;=}\nonumber\\ 
&& \frac{\erf(r/a)}{r} - \frac{\hbar^2}{\sqrt{\pi} \,m^2 a^5}
\exp\left(-\frac{r^2}{a^2}\right) \, t^2 + \order{t^4} \,.
\end{eqnarray}
Here, however, we choose an expansion around $a=\infty$, which is justified in all usual experimental
situations -- which is why one usually assumes plane-wave solutions. This approximation yields
\begin{equation}\begin{split} \label{eqn:width-expansion}
 f(t,r;m,a) = \frac{2}{\sqrt{\pi} \, a} \, \left( 1 - \frac{r^2}{3 a^2} + \frac{r^4}{10 a^4} \right)\\
-\frac{\hbar^2 \, t^2}{\sqrt{\pi}\, m^2 \, a^5}+ \order{a^{-7}} \,,\end{split}
\end{equation}
which is time-independent up to order $a^{-5}$.\\
\\
\textbf{Scenario 1: Unique spatial wave-function}\\
\\
In the case of equation \eqref{eqn:sn_one_wf_two_masses} we have
\begin{eqnarray} \label{eqn:sn_unique_wf}
&&\rmi \hbar \partial_t \psi\;=\; H_\text{kin}(m_K)\, \psi\\
&&\quad- G m_K \left(\abs{\alpha}^2 m_S+\abs{\beta}^2 m_L\right) \, f(t,r;m_K,a) \, \psi \,.\nonumber
\end{eqnarray}
If we then write $m_S = m$, $m_L = m + \Delta m$ and use the expansion \eqref{eqn:width-expansion} we get
\begin{eqnarray}
&&\rmi \hbar \partial_t \psi\; =\; H_\text{kin}(m_K) \psi \nonumber\\
&&-  \frac{2 G m_K}{\sqrt{\pi} \, a} \left(m_K+\abs{\beta(t)}^2 \Delta m\right) \, \psi \,.
\end{eqnarray}
This is time dependent only through the coefficient $\beta$. We also used \eqref{eqn:complex_eigenval} to
replace $m$ by $m_K$ in the approximation. \\
\\
\textbf{Scenario 2: Different spatial wave-function}\\
\\
If we assume two different wave-functions, as in equations~\eqref{eqn:sn-different-wavefunctions},
we can write them as
\begin{subequations}\label{eqn:sn_diff_wf}\begin{align}
 \rmi \hbar \partial_t\, \psi_S &= H_\text{kin}(m_S)\, \psi_S \nnl
&\bleq - G m_S^2\;  \abs{\alpha(t)}^2 \, f(t,r;m_S,a)\, \psi_S  \nnl
&\bleq - G m_S m_L\;  \abs{\beta(t)}^2 \, f(t,r;m_L,a)\, \psi_S \\
 \rmi \hbar \partial_t\, \psi_L &= H_\text{kin}(m_L)\, \psi_S \nnl
&\bleq - G m_S m_L\;  \abs{\alpha(t)}^2 \, f(t,r;m_S,a)\, \psi_L  \nnl
&\bleq - G m_L^2\;  \abs{\beta(t)}^2 \, f(t,r;m_L,a)\, \psi_L \,.
\end{align}\end{subequations}
If we then again write $m_S = m$, $m_L = m + \Delta m$ and use the expansion \eqref{eqn:width-expansion} we get
\begin{subequations}\begin{align}
 \rmi \hbar \partial_t \psi_S &= H_\text{kin}(m) \psi_S \nnl
&\bleq \hspace{-35pt} - \frac{2 G \, m^2}{\sqrt{\pi} \, a} \, \left[ 1
+ \left(\frac{1}{m} +\frac{\hbar^2 t^2}{m^3 a^4} \right) \, \abs{\beta(t)}^2 \, \Delta m \right] \, \psi_S \\
\rmi \hbar \partial_t \psi_L &= H_\text{kin}(m_L) \psi_L \nnl
&\bleq \hspace{-35pt} - \frac{2 G \, m_L^2}{\sqrt{\pi} \, a} \, \left[ 1
+ \left(\frac{1}{m} +\frac{\hbar^2 t^2}{m^3 a^4} \right) \, \abs{\beta(t)}^2 \, \Delta m \right] \, \psi_L
\end{align}\end{subequations}

\subsection{Gravity-induced energy shift}

Above we obtained the nonlinear \schr equation which describes the dynamics of the
kaon system in the presence of gravity, and therefore an approximation of the space-dependent Hamiltonian.
The Hamiltonian that governs the flavor oscillations is modified by the energy shift due to the gravitational
interaction. In order to calculate this energy shift, we must consider the expectation value of the Hamiltonian.
This expectation value is proportional to
\begin{equation}
 \langle \psi^f \vert f(t,r;m,a) \vert \psi^f \rangle
= \sqrt{\frac{2}{\pi \, a^2}} \, \left( 1 + \frac{\hbar^2 t^2}{m^2 a^4} \right)^{-1/2} \,,
\end{equation}
where we approximated the wave-function by the solution of the free \schr equation, as previously explained.
For the mass $m +\Delta m$ we can expand this and obtain up to first order in $\Delta m$:
\begin{equation}\begin{split}
 \langle \psi^f \vert f(t,r;m+\Delta m,a) \vert \psi^f \rangle
=  \langle \psi^f \vert f(t,r;m,a) \vert \psi^f \rangle \\
+ \frac{\hbar^2 t^2}{m^3 a^4} \, \left( 1 + \frac{\hbar^2 t^2}{m^2 a^4} \right)^{-3/2} \, \Delta m \,.
\end{split}\end{equation}
\\
\\
\textbf{Scenario 1: Unique spatial wave-function}\\
\\
From equation \eqref{eqn:sn_unique_wf} we get the energy shift
\begin{align}
 \Delta E &= - G m_K \left(\abs{\alpha}^2 m_S+\abs{\beta}^2 m_L\right) \nnl
&\bleq \times \langle \psi^f \vert f(t,r;m_K,a) \vert \psi^f \rangle \nnl
&= - \sqrt{\frac{2}{\pi}} \, \frac{G m_K}{a} \, \left(m+\abs{\beta}^2 \Delta m\right)
+ \order{a^{-6}} \,.
\end{align}
Both states $\ket{K_S}$ and $\ket{K_L}$ obtain a constant energy shift, but these only yield
a constant phase shift. The contribution proportional to $\Delta m$, however, adds to the flavor
oscillations~\eqref{eqn:flavor-oscillations}. It acts like a shift of the mass difference:
\begin{equation}
 \Delta m \to \left(1 -  \Delta_{SN}  \right) \,\Delta m
\end{equation}
with
\begin{equation}
 \Delta_{SN} =  \sqrt{\frac{2}{\pi}} \, \frac{G m_K}{c^2 a}\,.
\end{equation}
\\
\\
\textbf{Scenario 2: Different spatial wave-function}\\
\\
From equations \eqref{eqn:sn_diff_wf} one obtains
\begin{subequations}\begin{align}
\Delta E_S &= -\sqrt{\frac{2}{\pi}} \frac{G \, m}{a} \left[m +  \abs{\beta}^2 \,\Delta m \right]
+ \order{a^{-4}} \\
\Delta E_L &= -\sqrt{\frac{2}{\pi}} \frac{G \, m}{a} \left[m +  \left(1+\abs{\beta}^2\right)
\,\Delta m \right] + \order{a^{-4}} \,,
\end{align}\end{subequations}
where higher order terms in $1/a$ have been omitted. The shift in $\Delta m$ therefore is twice
the one before:
\begin{equation}
 \Delta m \to \left(1 - 2\cdot \Delta_{SN}  \right) \,\Delta m \,,
\end{equation}
where we assume that $m \approx m_K$. Inserting the kaon mass, one finds that a large effect is only
expected if the wave-function becomes close to or narrower than about $\unit{10^{-54}}{\meter}$, far below
the Planck length. This result does not change for other meson types.

Hence, we conclude that, although the effect is unobservably small for the kaon, the resulting effect
depends on the treatment of the spatial wave-function, either by a unique wave-function or by different
wave-functions for the different mass eigenstates.
Let us also remark here that the effect does depend on the particular shape of the spacial wave-function about which is not
known much from the experiments. Thus dedicated experiments
to study the spatial wave-function of mesons would help to
understand the right treatment for possible nonlinear modifications of
the Schrödinger equation.

In addition to the energy shift, the \sne of course also leads to the usual localization of the wave-function as it has
been described in~\cite{Giulini:2011}, which is a very small effect due to
the weakness of the gravitational self-interaction in the situation at hand (large wave-function, small masses).

\section{Comparison to the CSL collapse model}

Collapse models~\cite{Bassi:2013} predict the spontaneous collapse of the wave-function, in order to avoid the emergence
of macroscopic superpositions. In their mass-dependent formulation, they claim that the collapse of any system’s wave-
function depends on its mass. Recently, the most popular collapse model, the mass-proportional CSL (Continuous
Spontaneous Localization) model was applied to the meson-antimeson systems~\cite{Donadi:2013}. Here, the crucial point
was again to connect the spatial with the flavor wave-function part. The authors chose the sum of the kinetic
contribution of the short and long-lived component. After a cumbersome computation solving the stochastic nonlinear
differential equation they found the following probability
\begin{equation}\begin{split}\label{finalsingleprobability2}
P(K^0 t;|K^0|) = \frac{1}{4}\biggl( \rme^{-\Gamma_{S} t}+\rme^{-\Gamma_{L}t} \\
 +2\cos(\Delta m t)\, \rme^{-\Gamma t} \,
\underbrace{\rme^{-\frac{\gamma \Delta m^{2}}{16\pi^{3/2}r_{C}^{3}m_{0}^{2}} t}}_{\textrm{effect due to CLS model}}\biggr)
\end{split}\end{equation}
where the collapse rate $\gamma$ and the coherence length $r_C$ are parameters of the collapse model and $m_{0}$ is a
reference mass. Thus, in opposition to the solution of the \sne, the mass-dependent collapse effect leads to a damping
proportional to $\Delta m^2$ and has, consequently, to be compared to decoherence effects.

\section{Summary and conclusions}

The aim of this contribution was to investigate how Newtonian self-gravitation can be included in the standard framework
to handle flavor oscillations of neutral meson systems.
Provided that such a hypothetical nonlinear modification of the quantum dynamics due to gravity
would be correct, it is not straightforwardly physically intuitive which mass is the
relevant one for the different terms in the Hamiltonian. Moreover, for any nonlinear extension of the Schr\"odinger
evolution, as in the case of the \sne or spatial collapse models, one has to assume a certain relation between the spatial
and flavor wave-functions. We have considered two possible scenarios, a separable and entangled ansatz, and derived the
effect of the \sne under certain assumptions.
Although the effect turns out to be unobservably small, we find that there is a conceptually different
result for the two scenarios, namely a shift in energy which is twice the one for the scenario of a unique
wave-function of the two mass-energy eigenstates.
The correct treatment of the spatial wave-function in the quantum mechanical description of non-relativistic
elementary particle systems therefore is a crucial question, which deserves further consideration,
independently of the question if the kind of self-gravitation discussed in this paper actually exists.

Let us remark that the non-Hermitian part of the Hamiltonian was not considered or affected by the \sne. This is
consistent with the results in ref.~\cite{Bertlmann:2006} where the authors showed that the non-Hermitian part can be
removed by changing the Schr\"odinger equation into a Lindblad equation by extending the Hilbert space. Also the $CP$
violation in mixing was not affected by self-gravitation since it does not enter in the eigenvalues of the considered
Hamiltonian. It is not clear what happens with direct $CP$ violating effects, i.\,e. where the violation occurs on the
amplitude level.

Let us also mention that neutrinos exhibit a similar oscillation feature in flavor space, hence the same
conceptual problems apply to this case, although the relevant masses are even smaller.

Our current understanding of the non-relativistic limit of particle physics, as well as the foundations of the
\sne (provided it is correct), is not sufficient to derive the \schr--Newton effects unambiguously.
In particular, there is no definite answer, which are the relevant masses and how to treat the spatial wave-function.
However, none of the possibilities we discussed leads to any conceptional contradictions between
the \sne and the treatment of kaons as a non-relativistic  quantum system.
In all considered cases the effects within our assumptions were found to be irrelevant in practical situations,
therefore no contradictions to already performed or envisaged experiments in elementary particle physics seem to exist.

\begin{acknowledgments}
Both authors want to gratefully thank the COST action MP1006 ``\textit{Fundamental Problems
in Quantum Physics}'' initiating this research. AG acknowledges support from the John Templeton foundation
(grant 39530) and Deutsche Forschungsgemeinschaft. BCH gratefully acknowledges support from the Austrian Science Fund (FWF-P26783).
\end{acknowledgments}

\appendix
\section{Derivation of the \sne from a doublet-state formalism}\label{app:doublet}

In~\cite{Bahrami:2014} the \sne\ has been derived from semi-classical gravity for a scalar field.
In the non-relativistic limit, one ends up with the gravitational interaction Hamiltonian
\begin{equation}
\label{eqn:scalar-hint}
 \hat{H}_\text{int} = -G \int \D^3 r \, \D^3 r' \frac{\bra{\Psi}
\hat{\varrho}(\vec r')\ket{\Psi}}{\abs{\vec r - \vec r'}} \, \hat{\varrho}(\vec r) \,,
\end{equation}
where $\hat{\varrho} = m \hat{\psi}^\dagger \hat{\psi}$ is the mass density operator for only one kind of particles.
This, together with the kinetic energy operator
\begin{equation}
\label{eqn:scalar-kin-energy}
\hat{T} = \int \D^3 r \, \hat{\psi}^\dagger(\vec r) \left(-\frac{\hbar^2}{2m} \nabla^2 \right) \hat{\psi}(\vec r) \,,
\end{equation}
yields the \sne\ in second quantized formalism,
\begin{equation}\label{eqn:second-quantized-sn}
 \rmi \hbar \, \partial_t \, \ket{\Psi} = \left( \hat{T} + \hat{H}_\text{int} \right) \, \ket{\Psi} \,.
\end{equation}
The standard, first quantized \sne\ follows for the one-particle state
\begin{equation}
\ket{\Psi} = \int \D^3 r \, \Psi(t,\vec r) \, \hat{\psi}^\dagger(\vec r) \, \ket{0} \,,
\end{equation}
where $\Psi(t,\vec r)$ is the spatial wave-function.

Let us extend this to a doublet field, as in the case of the neutral kaon. Hence, we consider the one-particle states
\begin{equation}\label{eqn:meson-doublet-state}
\ket{\Psi} = \int \D^3 r \, \tvector{\Psi_1(t,\vec r) \, \hat{\psi_1}^\dagger(\vec r)}%
{\Psi_2(t,\vec r) \, \hat{\psi_2}^\dagger(\vec r)} \, \ket{0} \,,
\end{equation}
where the indices 1 and 2 denote a not yet specified orthonormal basis in the space of the field operators.
Note that this in principle covers the case where both fields evolve differently in space, as well as the case where
both have the same unique distribution in space. The latter case is obtained by demanding $\Psi_1 \equiv \Psi_2$.
The normalization condition $\langle \Psi \vert \Psi \rangle = 1$ requires $\abs{\Psi_1}^2 + \abs{\Psi_2}^2 = 1$.

We will not specify the form of the kinetic energy and mass density operators, yet, but make the general ansatz
\begin{align}
\hat{T} &= \int \D^3 r \, \tvector{\hat{\psi_1}(\vec r)}{\hat{\psi_2}(\vec r)}^\dagger \,
\tmatrix{\hat{T}_{11}}{\hat{T}_{12}}{\hat{T}_{21}}{\hat{T}_{22}} \,
\tvector{\hat{\psi_1}(\vec r)}{\hat{\psi_2}(\vec r)} \\
\hat{\rho}(\vec r) &= \tvector{\hat{\psi_1}(\vec r)}{\hat{\psi_2}(\vec r)}^\dagger \,
\tmatrix{M_{11}}{M_{12}}{M_{21}}{M_{22}} \,
\tvector{\hat{\psi_1}(\vec r)}{\hat{\psi_2}(\vec r)} \,,
\end{align}
where $\hat{T}_{ij}$ are operators acting on the spatial wave-function.
Applied to the state~\eqref{eqn:meson-doublet-state} these yield
\begin{align}
\hat{T} \ket{\Psi} &= \int \D^3 r \,
\tvector{\hat{T}_{11} \Psi_1(t,\vec r) \hat{\psi}_1^\dagger + \hat{T}_{21} \Psi_1(t,\vec r) \hat{\psi}_2^\dagger}%
{\hat{T}_{12} \Psi_2(t,\vec r) \hat{\psi}_1^\dagger + \hat{T}_{22} \Psi_2(t,\vec r) \hat{\psi}_2^\dagger} \, \ket{0} \\
\hat{\rho}(\vec r) \ket{\Psi} &=
\tvector{M_{11} \Psi_1(t,\vec r) \hat{\psi}_1^\dagger + M_{21} \Psi_1(t,\vec r) \hat{\psi}_2^\dagger}%
{M_{12} \Psi_2(t,\vec r) \hat{\psi}_1^\dagger + M_{22} \Psi_2(t,\vec r) \hat{\psi}_2^\dagger} \, \ket{0} \,,
\end{align}
and the expectation value of the mass density operator is
\begin{equation}
\bra{\Psi} \hat{\rho}(\vec r) \ket{\Psi}_t = M_{11} \abs{\Psi_1(t,\vec r)}^2 + M_{22} \abs{\Psi_2(t,\vec r)}^2 \,.
\end{equation}
By inserting these expressions into the second quantized \sne~\eqref{eqn:second-quantized-sn}, and multiplying with
$\bra{0} \hat{\psi}_{1,2}$ from the left, one obtains the following coupled system of equations:
\begin{widetext}
\begin{subequations}\begin{align}
\label{eqn:first-sn-eq}
\rmi \hbar \, \partial_t \, \Psi_1(t,\vec r) &= \hat{T}_{11} \, \Psi_1(t,\vec r) \nnl
&\bleq - G \, M_{11} \, \int \D^3 r' \, \left( M_{11} \, \frac{\abs{\Psi_1(t,\vec r')}^2}{\abs{\vec r - \vec r'}}
+ M_{22} \, \frac{\abs{\Psi_2(t,\vec r')}^2}{\abs{\vec r - \vec r'}} \right) \, \Psi_1(t,\vec r) \\
\label{eqn:second-sn-eq}
\rmi \hbar \, \partial_t \, \Psi_2(t,\vec r) &= \hat{T}_{22} \, \Psi_2(t,\vec r) \nnl
&\bleq - G \, M_{22} \, \int \D^3 r' \, \left( M_{11} \, \frac{\abs{\Psi_1(t,\vec r')}^2}{\abs{\vec r - \vec r'}}
+ M_{22} \, \frac{\abs{\Psi_2(t,\vec r')}^2}{\abs{\vec r - \vec r'}} \right) \, \Psi_2(t,\vec r) \\
0 &= \hat{T}_{21} \, \Psi_1(t,\vec r) \nnl
&\bleq - G \, M_{21} \, \int \D^3 r' \, \left( M_{11} \, \frac{\abs{\Psi_1(t,\vec r')}^2}{\abs{\vec r - \vec r'}}
+ M_{22} \, \frac{\abs{\Psi_2(t,\vec r')}^2}{\abs{\vec r - \vec r'}} \right) \, \Psi_1(t,\vec r) \\
0 &= \hat{T}_{12} \, \Psi_2(t,\vec r) \nnl
&\bleq - G \, M_{12} \, \int \D^3 r' \, \left( M_{11} \, \frac{\abs{\Psi_1(t,\vec r')}^2}{\abs{\vec r - \vec r'}}
+ M_{22} \, \frac{\abs{\Psi_2(t,\vec r')}^2}{\abs{\vec r - \vec r'}} \right) \, \Psi_2(t,\vec r) \,.
\end{align}\end{subequations}
\end{widetext}
In the case of free evolution, without gravity, the last two equations require the off-diagonal terms of
the kinetic energy operator to vanish. The first two equations are completely decoupled in this case.
Therefore, at this point, no restriction on the relation of the two wave-functions can be made. In particular,
it is possible that $\Psi_1 \equiv \Psi_2$, which then implies $\hat{T}_{11} = \hat{T}_{22}$.

However, if the \schr--Newton term is considered, the off-diagonal elements of the kinetic energy operator vanish
if and only if the mass matrix is diagonal. The self-gravitational interaction also leads to a coupling of
equations~\eqref{eqn:first-sn-eq} and~\eqref{eqn:second-sn-eq}. Having $\Psi_1 \equiv \Psi_2$ then not only
requires $\hat{T}_{11} = \hat{T}_{22}$ but also $M_{11} = M_{22}$.

Hence, if there is a mass difference of the states, as for the neutral kaon system, a doublet-state formalism is
only compatible with the \sne if both states evolve in with their respective masses, having different wave-functions
$\Psi_1 \not= \Psi_2$.

\end{document}